\documentclass[showpacs,preprintnumbers,amsmath,amssymb,prb]{revtex4}

\usepackage{graphicx}
\usepackage{dcolumn}
\usepackage{bm}

\begin{document}

\title{Antiferromagnets with random vacancies and substitutional spins on the triangular lattice}

\author{A.\ V.\ Syromyatnikov}
\email{asyromyatnikov@yandex.ru}
\author{F.\ D.\ Timkovskii}
\email{philippinho@yandex.ru}
\affiliation{National Research Center "Kurchatov Institute" B.P.\ Konstantinov Petersburg Nuclear Physics Institute, Gatchina 188300, Russia}

\date{\today}

\begin{abstract}

We discuss theoretically static and dynamical properties of $XY$ and Heisenberg antiferromagnets on the triangular lattice with random vacancies and substitutional spins. It is shown that the distortion of $120^\circ$ magnetic order produced by a single defect is described by electrostatic equations for a field of an electrically neutral complex of six charges located around the impurity. The first finite term in the multipole expansion of this field is the octupole moment which decays as $1/r^3$ with the distance $r$. The linearity of equations allows to describe analytically the distortion of the long-range magnetic order at a small concentration $c$ of defects. We obtain analytically renormalization of the elastic neutron scattering cross section and the magnon spectrum $\epsilon_{\bf k}$ in the leading order in $c$. We find that the scattering on impurities renormalizes weakly the bare spectrum $\epsilon_{\bf k}\propto k$ at $k\gg\sqrt c$. However the renormalization is substantial of the long-wavelength magnon spectrum at $k\ll\sqrt c$: $\epsilon_{\bf k}\propto \sqrt{c /\ln(1/k)}$ at $k\to0$ and there is a parametrically large region in which magnons with not too small momenta are overdamped and localized. This strong modification of the long-wavelength spectrum leads to the stabilization of the slightly distorted magnetic long-range order at $T<T_N\sim S^2J/\ln(1/c)$ and to the considerable change in the density of states and in the specific heat. The overdamped modes arise also in quasi-2D spin systems on a stacked triangular lattice. 

\end{abstract}

\pacs{64.70.Tg, 72.15.Rn, 74.40.Kb}

\maketitle

\section{Introduction}

The influence of quenched disorder on strongly correlated electron systems remains one of the hottest topics in modern condensed matter physics. Even a small amount of disorder can change drastically properties of some systems, the most famous examples being the Anderson localization, the Kondo effect, order-by-disorder phenomena in frustrated spin systems, and disorder-induced new phases and new types of phase transitions.

The influence of a small concentration of point defects (vacancies, substitutional or interstitial spins) on magnets with a collinear spin ordering has been understood quite well using, in particular, the so-called $T$-matrix formalism which allows to calculate observable quantities in the first order in the defect concentration. \cite{izyum,orderdif,orderdif2,wan,2dvac} In many cases, the introduction of a defect does not perturb the local magnetic ordering in the host system that makes straightforward the application of the Holstein-Primakoff formalism and the $T$-matrix method. In 3D magnets with Goldstone excitations, a small concentration $c\ll1$ of impurities produces just a small renormalization of the spin-wave velocity, the spin stiffness, the magnetic susceptibility and leads to a small magnon damping which is normally much smaller than the hydrodynamic damping \cite{hydro,hydro2}. The renormalization of the spin-wave spectrum is known to be much stronger in 2D collinear antiferromagnets (AFs). \cite{wan,2dvac,syromyat1,syromyat2} In particular, magnons with wavelength $\lambda\agt e^{\pi/4c}$ were found to be overdamped and localized in 2D Heisenberg AF on depleted square lattice \cite{2dvac} while the long-range order persists in this system up to the percolation threshold \cite{lro2d1,lro2d2}.

Point defects are much less studied in non-collinear spin systems mainly because any impurity disturbs the local magnetic order by producing a long-range spin texture. \cite{brenig,afh1,wollny1,wollny2,olegdm} It was difficult to apply the well-developed $T$-matrix formalism to non-collinear systems with defects because it was unclear how to take into account the long-range textures from different randomly distributed impurities and their interference. For instance, investigation \cite{brenig} of dynamical properties of 2D Heisenberg AF in magnetic field using the $T$-matrix technique relies on numerical results for spin textures.

It is quite an old but seemingly well forgotten idea by Villain \cite{vill1,vill2} that the impurity-induced spin textures in Heisenberg magnets is described by electrostatic equations. This idea was particularly useful in considerations of frustrated defects in Heisenberg AF on simple square lattice \cite{aharony,korenblit} and in recent study \cite{olegdm} of 3D spiral magnets. It was shown in Ref.~\cite{olegdm} that the distortion of the spiral order arisen around a couple of spins whose exchange coupling constant is changed (a defect bond) is given by the field of an electric dipole. Due to the linearity of equations describing the perturbation of the long-range order by impurities, the solution of the many-defect problem was a simple superposition of single-impurity solutions (the superposition principle in electrostatics). This helped to incorporate the $T$-matrix formalism into the spin-wave theory of the non-collinear magnets, to perform the averaging over disorder configurations, and to consider analytically the renormalized spin dynamics. \cite{olegdm}

\begin{figure}
\includegraphics[scale=0.9]{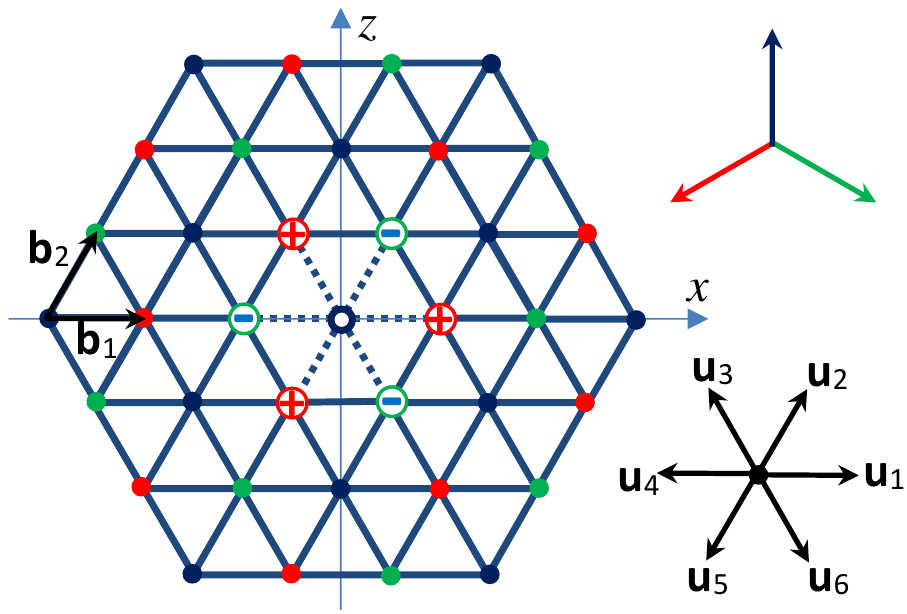}
\caption{Antiferromagnet on the triangular lattice with a defect considered in the present paper. Bold and dashed lines denote exchange couplings with constants $J$ and $\cal J$, respectively. The vacancy (a free spin) corresponds to ${\cal J}=0$.  We consider also the substitutional spin at the center of the defect whose value $S'$ differs from the spin value in the host system $S$. Sites belonging to different sublattices of the bare $120^\circ$ magnetic order are distinguished by color. The distortion of the magnetic order away from the impurity is described by an equation for the field of electric charges shown around the impurity whose values are proportional to $J-{\cal J}S'/S$. Vectors ${\bf u}_i$ connect a lattice site with its six neighbors. Lattice translation vectors ${\bf b}_{1,2}$ are also shown.
\label{lattice}}
\end{figure}

In the present paper, we apply the method proposed in Ref.~\cite{olegdm} to discuss vacancies and substitutional defects in classical Heisenberg and $XY$ AFs on the triangular lattice. It is natural to assume that the substitution of a magnetic ion by another one changes in a real material also the exchange coupling interactions of this ion with its nearest neighbors. Then, we consider below the model shown in Fig.~\ref{lattice}. The vacancy corresponds in this scheme to the zero exchange coupling constant of the substitutional spin with its neighbors (a free spin). Importantly, we consider a symmetric defect in which six couplings with six neighbors are the same. As it is shown in Ref.~\cite{olegdm}, the distortion from a single defect bond is described by the field of the electric dipole which decreases in $d$-dimensional system as $1/r^{d-1}$ with the distance $r$. Such single-bond disorder destroys the long-range ordering in non-collinear systems at $d\le2$ and finite $c$ (see also below). \cite{wollny3} In contrast, the perturbation decays faster ($\sim1/r^{d+1}$) from a defect containing a symmetric complex of defect bonds the sum of dipolar moments of which is zero (such as that shown in Fig.~\ref{lattice}). As we show below, a finite concentration of considered impurities disturbs weakly the $120^\circ$ magnetic order.

The rest of the present paper is organized as follows. We discuss in Sec.~\ref{pure} the Heisenberg AF on the triangular lattice without defect and introduce the formalism which is used in further analysis. The distortion of the long-range magnetic order by a single defect is considered in Sec.~\ref{order}. We derive an analytic expression describing the spin texture produced by an isolated impurity of the considered type. In particular case of a vacancy, our general analytical results are in good quantitative agreement with previous numerical simulations \cite{wollny1}. Elastic neutron scattering cross section is also derived in Sec.~\ref{order} at finite concentration of impurities. We show that in contrast to the single-bond disorder in 3D spiral magnets, where the diffuse magnetic scattering has the form of power-law decaying tails of each Bragg peak \cite{olegdm}, the symmetric defects produce much less pronounced diffuse cross section.

We discuss renormalization of the magnon spectrum caused by the scattering on defects using the self-consistent $T$-matrix approach in Secs.~\ref{magnon} and \ref{subst}. To light notations, we separate the cases of $S'=S$ (Sec.~\ref{magnon}) and $S'\ne S$ (Secs.~\ref{subst}), where $S'$ and $S$ are values of substitutional spins and spins in the host system, respectively. At all model parameters, we find that the bare spectrum $\epsilon_{\bf k}\propto k$ is renormalized weakly at $k\gg\sqrt c$. In contrast, the spectrum changes drastically of magnons whose wavelengths is larger than the mean distance between impurities $1/\sqrt c$. We observe a well-defined quasiparticles at $k\to0$ with the spectrum $\epsilon_{\bf k}\propto \sqrt{c /\ln(1/k)}$. A parametrically large region is found in which magnons with not too small momenta are overdamped and localized. Such strong spectrum renormalization should be contrasted with much more weaker effects found before in collinear 2D spin systems. Notice also that a strong magnon energy renormalization was obtained also at $k\ll\sqrt c$ in 3D spiral magnets in Ref.~\cite{olegdm} while the defect-induced damping was found to be small.

We consider changes of static properties in Sec.~\ref{static}. The order parameter is discussed there and it is shown that the strong renormalization of the long-wavelength spectrum leads to the stabilization of the slightly distorted $120^\circ$ order at $T<T_N\sim S^2J/\ln(1/c)$ (the order-by-disorder effect). Strong low-energy modification of the density of states is also observed which can be visible in the specific heat temperature dependence. Notice that all particular derivations are performed below for the Heisenberg AF. We introduce the $XY$ anisotropy to the system in Sec.~\ref{disc} and argue that many results obtained in the Heisenberg AF remain valid in this case. The effect is also considered briefly in Sec.~\ref{disc} of small low-symmetry spin interactions and/or non-frustrated inter-plane exchange coupling in the stacked triangular lattice. Sec.~\ref{conc} contains our conclusion.

\section{Pure system}
\label{pure}

The Hamiltonian of the pure Heisenberg AF on the triangular lattice has the form
\begin{equation}
  \label{ham}
  \mathcal{H} = \sum_{\langle i,j \rangle} {\bf S}_{{\bf r}_i} {\bf S}_{{\bf r}_j},
\end{equation}
where the exchange coupling constant $J=1$ and $\langle ij\rangle$ denote nearest neighbor sites. Spin components in the laboratory coordinate frame $(x,y,z)$ are related as follows with spin components in the local frame $(x',y',z')$ at site $j$ in which $z'$ axis is directed along the local magnetization:
\begin{eqnarray}
  \label{transform}
	S^{x}_{{\bf r}_j} &=& S^{x^\prime}_{{\bf r}_j} \cos({\bf k}_0{\bf r}_j) + S^{z^\prime}_{{\bf r}_j} \sin({\bf k}_0{{\bf r}_j}),\nonumber\\
  S^{y}_{{\bf r}_j} &=& S^{y^\prime}_{{\bf r}_j} ,\\
	S^{z}_{{\bf r}_j} &=& -S^{x^\prime}_{{\bf r}_j} \sin({\bf k}_0{\bf r}_j) + S^{z^\prime}_{{\bf r}_j} \cos({\bf k}_0{\bf r}_j),\nonumber
\end{eqnarray}
where we assume for simplicity that spins lie within $xz$ plane and ${\bf k}_0 = \frac{4\pi}{3}{\bf b}_1$ is the vector of the magnetic structure which describes the 120$^\circ$ magnetic oder (see Fig.~\ref{lattice}).
\footnote{
There are two possible values for ${\bf k}_0$, $\frac{4\pi}{3}{\bf b}_1$ and $-\frac{4\pi}{3}{\bf b}_1$, where ${\bf b}_{1}$ is the lattice translation vector (see Fig.~\ref{lattice}), which describe 120$^\circ$ magnetic orders with one of two possible chiral orderings (see, e.g., Ref.~\cite{TAF}). We assume in this paper for definiteness that ${\bf k}_0 = \frac{4\pi}{3}{\bf b}_1$.
}
We use below the Holstein-Primakoff spin representation
\begin{eqnarray}
  \label{spinrep}
	S^{x^{\prime}}_{{\bf r}_j}
	&\approx&
	\sqrt{\frac{S}{2}}\left(a^\dagger_{{\bf r}_j}+a_{{\bf r}_j}-\frac{a^\dagger_{{\bf r}_j} a^2_{{\bf r}_j}}{4S}- \frac{a^{\dagger2}_{{\bf r}_j} a^{}_{{\bf r}_j}}{4S}\right), \nonumber \\
  S^{y^{\prime}}_{{\bf r}_j}
	&\approx& 
	i\sqrt{\frac{S}{2}}\left(a^\dagger_{{\bf r}_j}-a_{{\bf r}_j}+\frac{a^\dagger_{{\bf r}_j} a^2_{{\bf r}_j}}{4S}-\frac{a^{\dagger2}_{{\bf r}_j} a^{}_{{\bf r}_j}}{4S}\right),\\
	S^{z^{\prime}}_{{\bf r}_j} &=& S-a^\dagger_{{\bf r}_j} a^{}_{{\bf r}_j}.  \nonumber
\end{eqnarray}
After substitution of Eqs.~\eqref{transform} and \eqref{spinrep} to Eq.~\eqref{ham}, one obtains that there are no terms in the Hamiltonian linear in Bose operators and terms containing products of two and three operators have the form
\begin{eqnarray}
\label{h2}
	\mathcal{H}_2 &=& 
	\frac S8 \sum_i\sum_{j=1}^6 
	\left( 4a^\dagger_{{\bf r}_i} a^{}_{{\bf r}_i} + a^\dagger_{{\bf r}_i} a^{}_{{\bf r}_i+{\bf u}_j} + a^\dagger_{{\bf r}_i+{\bf u}_j} a^{}_{{\bf r}_i} - 3a^\dagger_{{\bf r}_i} a^\dagger_{{\bf r}_i+{\bf u}_j} - 3a^{}_{{\bf r}_i} a^{}_{{\bf r}_i+{\bf u}_j} \right),\\
\label{h3}
	\mathcal{H}_3 &=& 
	-\sqrt{\frac S2} \sum_i\sum_{j=1}^6 
	\sin({\bf k}_0{\bf u}_j)
	\left( a^\dagger_{{\bf r}_i} + a^{}_{{\bf r}_i} \right) a^\dagger_{{\bf r}_i+{\bf u}_j} a^{}_{{\bf r}_i+{\bf u}_j},
\end{eqnarray}
correspondingly, where ${\bf u}_j$ are vectors connecting a lattice site with its six neighbors (see Fig.~\ref{lattice}).

One has from Eq.~\eqref{h2} after the Fourier transform
\begin{eqnarray}
\label{h2f}
{\cal H}_2 &=& \sum_{\bf k} 
\left[
E_{\bf k} a^\dagger_{\bf k}a_{\bf k} 
+ 
\frac{B_{\bf k}}{2} \left(a_{\bf k}a_{-\bf k} + a^\dagger_{\bf k}a^\dagger_{-\bf k}\right)
\right],
\end{eqnarray}
where $E_{\bf k}=S(3+\lambda_{\bf k}/2)$, $B_{\bf k}=-S3\lambda_{\bf k}/2$, $\lambda_{\bf k} = \cos k_1+\cos k_2+\cos(k_1- k_2)$, ${\bf k} = k_1{\bf f}_1 + k_2{\bf f}_2$, and ${\bf f}_{1,2}$ are translation vectors of the reciprocal lattice. Introducing Green's functions $G(k) = \langle a_{\bf k}, a^\dagger_{\bf k} \rangle_\omega$, $F(k) = \langle a_{\bf k}, a_{-\bf k} \rangle_\omega$, ${\overline G}(k) = \langle a^\dagger_{-\bf k}, a_{-\bf k} \rangle_\omega$, and $F^\dagger (k) = \langle a^\dagger_{-\bf k}, a^\dagger_{\bf k} \rangle_\omega$ we have two sets of Dyson equations for them one of which has the form
\begin{equation}
\label{eqfunc}
\begin{aligned}
G(k) &= G^{(0)}(k) + G^{(0)}(k){\overline \Sigma}(k)G(k) + G^{(0)}(k) [B_{\bf k} + \Pi(k)] F^\dagger(k),\\
F^\dagger(k) &= {\overline G}^{(0)}(k) \Sigma(k)F^\dagger(k) + {\overline G}^{(0)}(k) [B_{\bf k} + \Pi^\dagger(k) ]G(k),
\end{aligned}
\end{equation}
where $k=(\omega,{\bf k})$, $G^{(0)}(k) = (\omega - E_{\bf k})^{-1}$ is the bare Green's function, and $\Sigma(k)$, $\overline \Sigma(k)$, $\Pi(k)$ and $\Pi^\dagger(k)$ are self-energy parts (see, e.g., Ref.~\cite{Syromyatnikov_2010}). One obtains solving Eqs.~(\ref{eqfunc}) and similar set of equations for $\overline G(k)$ and $F(k)$
\begin{equation}
\label{gf}
\begin{aligned}
&G(\omega,{\bf k}) = \frac{\omega + E_{\bf k} + \Sigma(\omega,{\bf k})}{{\cal D}(\omega,{\bf k})}, \qquad
& \overline G(\omega,{\bf k}) = \frac{-\omega + E_{\bf k} + \overline \Sigma(\omega,{\bf k})}{{\cal D}(\omega,{\bf k})},\\
&F(\omega,{\bf k}) = -\frac{B_{\bf k} + \Pi(\omega,{\bf k})}{{\cal D}(\omega,{\bf k})}, \qquad
&F^\dagger(\omega,{\bf k}) = -\frac{B_{\bf k} + \Pi^\dagger(\omega,{\bf k})}{{\cal D}(\omega,{\bf k})},
\end{aligned}
\end{equation}
where
\begin{eqnarray}
\label{d}
{\cal D}(\omega,{\bf k}) &=& \omega^2 - \left(\epsilon_{\bf k}^{(0)}\right)^2 - \Omega(\omega,{\bf k}),\\
\label{spec0}
\epsilon_{\bf k}^{(0)} &=& \sqrt{E_{\bf k}^2 - B_{\bf k}^2} = S\sqrt{(3-\lambda_{\bf k})(3+2\lambda_{\bf k})},\\
\label{o}
\Omega(\omega,{\bf k}) &=& E_{\bf k}\left(\Sigma + \overline{\Sigma}\right) 
- 
B_{\bf k} \left( \Pi + \Pi^\dagger \right) 
- 
\omega \left( \Sigma - \overline{\Sigma} \right) - \Pi\Pi^\dagger + \Sigma \overline{\Sigma},
\end{eqnarray}
and $\epsilon_{\bf k}^{(0)}$ is the spin-wave spectrum in the linear spin-wave approximation which is linear at $k\ll1$ and ${\bf k}\approx \pm{\bf k}_0$: 
\begin{equation}
\label{spec0l}
\epsilon_{\bf k}^{(0)}\approx \left\{
	\begin{aligned}
	& v_\| k, \quad & k\ll1,\\
	& v_\perp |{\bf k} \pm {\bf k}_0|, \quad & |{\bf k} \pm {\bf k}_0|\ll1,
	\end{aligned}
	\right.
\end{equation}
where
\begin{equation}
\label{vs}
		v_\| = \frac{3\sqrt3}{2}S, \qquad
		v_\perp = \frac{3\sqrt3}{2\sqrt2}S
\end{equation}
are spin-wave velocities in, respectively, longitudinal and transverse channels \cite{andreev,nlsmtriang,chubtri} (with respect to the plane in which spins lie). Quantity $\Omega(\omega,{\bf k})$ given by Eq.~(\ref{o}) describes the renormalization of the spin-wave spectrum square. We find $\Omega(\omega,{\bf k})$ in the leading order in the defect concentration in Sec.~\ref{magnon} within the $T$-matrix formalism.

\section{Distortion of the magnetic order by defects}
\label{order}

\subsection{Single impurity}

The Hamiltonian of the system with the impurity in the coordinate origin reads as
\begin{equation}
\label{ham2}
  \mathcal{H}_d = \mathcal{H} + \mathcal{V},
\end{equation}
where the first term is given by Eq.~\eqref{ham} and the second term describes the defect and has the form
\begin{equation}
  \label{vpert}
	\mathcal{V} = {\cal J} \sum_{j=1}^6 {\bf S}'_{\bf 0} {\bf S}_{{\bf u}_j}
	-
	 \sum_{j=1}^6 {\bf S}_{\bf 0} {\bf S}_{{\bf u}_j}, 
\end{equation}
where ${\bf u}_j$ are vectors connecting the center of the defect with neighboring sites (see Fig.~\ref{lattice}) and we assume that the spin value at the center of the defect $S'$ can differ from $S$. Eq.~\eqref{vpert} describes a vacancy (a free spin in the coordinate origin) at ${\cal J} =0$. 

One obtains from Eq.~\eqref{vpert} using Eqs.~\eqref{transform} and \eqref{spinrep} the following terms containing one, two, and three Bose operators:
\begin{eqnarray}
\label{v1}
  \mathcal{V}_1 &=& - U\frac{S^{3/2}}{\sqrt2}
	\sum_{j=1}^6 \sin({\bf k}_0{{\bf u}_j})
	\left( a^\dagger_{{\bf u}_j} + a^{}_{{\bf u}_j} \right),\\
\label{v2}
	\mathcal{V}_2 &=&
	\frac{S}{4}\sum_{j=1}^6
	\left( 
	2({\cal J}-1)a^\dagger_{\bf 0} a^{}_{\bf 0} + 2Ua^\dagger_{{\bf u}_j}a^{}_{{\bf u}_j} 
	+
	\left( {\cal J}\sqrt\beta-1 \right)
	\left( a^\dagger_{\bf 0}a^{}_{{\bf u}_j} +a^\dagger_{{\bf u}_j}a^{}_{\bf 0} -3a^\dagger_{{\bf u}_j}a^\dagger_{\bf 0} -3a^{}_{{\bf u}_j}a^{}_{\bf 0} \right)
	\right),\\
\label{v3}
		\mathcal{V}_3 &=&
	\frac 14 \sqrt{\frac S2} \sum_{j=1}^6 \sin({\bf k}_0{\bf u}_j)
	\left( 
	U a^\dagger_{{\bf u}_j}a^\dagger_{{\bf u}_j}a^{}_{{\bf u}_j} + U a^\dagger_{{\bf u}_j}a^{}_{{\bf u}_j}a^{}_{{\bf u}_j} 
	- 
	4 \left( {\cal J}\sqrt\beta-1 \right) (a^\dagger_{\bf 0}+a^{}_{\bf 0})a^\dagger_{{\bf u}_j}a^{}_{{\bf u}_j} 
\right.\nonumber\\
	&&\left.
	{}+ 
	4 ({\cal J}-1) (a^\dagger_{{\bf u}_j}+a^{}_{{\bf u}_j})a^\dagger_{\bf 0}a^{}_{\bf 0} 
	\right),
\end{eqnarray}
respectively, where 
\begin{eqnarray}
\label{u}
	U &=& {\cal J}\frac{S'}{S} - 1,\\
\label{beta}
	\beta &=& \frac{S'}{S}.
\end{eqnarray}
It will be clear soon that $U$ measures the defect "strength" (notice that it is the parameter which arose also in previous discussions \cite{izyum,orderdif,orderdif2,wan} of disordered collinear magnets). The appearance of terms \eqref{v1} in the Hamiltonian which are linear in Bose operators signifies the distortion of the bare 120$^\circ$ ordering. To eliminate the linear terms in the Hamiltonian, one has to make the shift
\begin{equation}
  \label{shift}
\begin{aligned}
	a_{{\bf r}_j} &\mapsto a_{{\bf r}_j}+\rho_{{\bf r}_j} e^{i \varphi_{{\bf r}_j}}, \\
  a^\dagger_{{\bf r}_j} &\mapsto a^\dagger_{{\bf r}_j}+\rho_{{\bf r}_j} e^{-i \varphi_{{\bf r}_j}} 
\end{aligned}
\end{equation}
throughout the system, where $\rho_{{\bf r}_j}$ and $\varphi_{{\bf r}_j}$ are real parameters which describe the perturbation of the spin ordering due to the defect. Both previous numerical consideration \cite{wollny1} and our accurate analysis show that all spins remain to lie within the $xz$ plane in the ground state after the introduction of the defect. 
Then, we put 
\begin{equation}
\label{phi}
	\varphi_{{\bf r}_j}=0
\end{equation}
 in which case the constant term vanishes in $S^y_{{\bf r}_j}$ after shift \eqref{shift} (see Eqs.~\eqref{transform} and \eqref{spinrep}). As it is seen from Eq.~\eqref{spinrep}, finite $\rho_{{\bf r}_j}$ describe a rotation of the magnetization at site ${\bf r}_j$ within $xz$ plane, the rotation angle being
 \begin{equation}
\label{theta}
\theta_{{\bf r}_j}	\approx \rho_{{\bf r}_j} \sqrt{\frac 2S}
 \end{equation}
because $S_{{\bf r}_j}^{x\prime}\approx\sqrt{2S} \rho_{{\bf r}_j}$ and $S_{{\bf r}_j}^{z\prime}=S-\rho_{{\bf r}_j}^2 \approx S$. We assume here and in the subsequent analysis that 
 \begin{equation}
\label{assum}
	|\rho_{{\bf r}_j}|\ll\sqrt{S}
 \end{equation}
which is justified below at $|U|\alt1$ by particular quantitative estimations and by comparison of our results with previous numerical simulations. Eq.~\eqref{assum} allows to restrict ourselves to terms of leading orders in $\rho_{{\bf r}_j}$ in further discussion.

Hamiltonian \eqref{ham2} acquires the form after shift \eqref{shift}
\begin{equation}
\label{hd}
  \mathcal{H}_{d} = \mathcal{E} + \mathcal{H}_{d1}+\mathcal{H}_{d2}+\dots,
\end{equation}
where $\mathcal{E}$, $\mathcal{H}_{d1}$, and $\mathcal{H}_{d2}$ stand for terms containing zero, one, and two Bose operators, respectively. In particular, one has
\begin{eqnarray}
	\label{h12}
  \mathcal{H}_{d1} 
	&=& 
	\sum_i a^\dagger_{{\bf r}_i} \frac S4 \sum_{j=1}^6 
	\left(2 \rho_{{\bf r}_i}-\rho_{{\bf r}_i+{\bf u}_j}- \rho_{{\bf r}_i-{\bf u}_j} \right)
	- 
	U\frac{S^{3/2}}{\sqrt2} \sum_{j=1}^6 
	a^\dagger_{{\bf u}_j} \sin({\bf k}_0{\bf u}_j)\nonumber\\
	&&{}- 
	\frac S2 \sum_{j=1}^6 
	\left(
	a^\dagger_{\bf 0} 
	\left( \left({\cal J}\sqrt\beta-1 \right) \rho_{{\bf u}_j} - \left({\cal J}-1\right)\rho_{\bf 0} \right)  
	- a^\dagger_{{\bf u}_j}
	\left( U\rho_{{\bf u}_j} - \left({\cal J}\sqrt\beta-1 \right) \rho_{\bf 0} \right) 
	\right)
	+ {\rm H.c.},
\end{eqnarray}
where the first, the second, and the third terms stem from Eqs.~\eqref{h2}, \eqref{v1}, and \eqref{v2}, respectively, and we take into account only terms ${\cal O}(\rho)$. Linear terms \eqref{h12} disappear in the Hamiltonian if the following equations hold at each site $i$:
\begin{eqnarray}
  \label{sys}
	\sum_{j=1}^6   
	\left(2 \rho_{{\bf r}_i}-\rho_{{\bf r}_i+{\bf u}_j}- \rho_{{\bf r}_i-{\bf u}_j} \right)
	&=& 
	2\sum_{j=1}^6  
	\left(
	\delta_{{{\bf r}_i},{\bf 0}}
	\left( \left({\cal J}\sqrt\beta-1 \right) \rho_{{\bf u}_j} - \left({\cal J}-1\right)\rho_{\bf 0} \right)  
	- \delta_{{{\bf r}_i},{{\bf u}_j}}
	\left( U\rho_{{\bf u}_j} - \left({\cal J}\sqrt\beta-1 \right) \rho_{\bf 0} \right)
	\right)\nonumber\\
	&&{}+ 
	2\sqrt{2S}U \sum_{j=1}^6 
	\delta_{{{\bf r}_i},{{\bf u}_j}} \sin({\bf k}_0{\bf u}_j),
	\qquad
	\forall i
\end{eqnarray}
where $\delta_{i,j}$ is the Kronecker delta. Minimization of the classical ground-state energy $\mathcal{E}$ in Eq.~\eqref{hd} with respect to $\rho_{{\bf r}_i}$ also leads to Eqs.~\eqref{sys}.

Notice that Eqs.~\eqref{sys} do not change after the replacement $\rho_{{\bf r}_i}\mapsto \rho_{{\bf r}_i}+C$ for all $i\ne0$ and $\rho_{\bf 0}\mapsto \rho_{\bf 0}+C\sqrt\beta$, where $C$ is a constant. This fact is related to the invariance of the system with respect to a global rotation of all spins by the same angle (see Eq.~\eqref{theta}). Then, we put $\rho_{\bf 0}=0$ below.

The expression in the left-hand side of Eq.~\eqref{sys} contains three discreet derivatives along three high-symmetry lattice directions. This allows to represent Eqs.~\eqref{sys} in the continuum limit as follows:
\begin{eqnarray}
\label{contlim}
  \Delta\rho({\bf r})
	&=&
	U\sqrt{\frac{2S}{3}} \sum_{j=1}^3
	\left[
	\left(\sqrt{\frac{2}{3S}}\rho_{{\bf u}_{2j-1}}+1\right) \delta\left({\bf r}-{\bf u}_{2j-1}\right)
	+
	\left(\sqrt{\frac{2}{3S}}\rho_{{\bf u}_{2j}}-1\right) \delta\left({\bf r}-{\bf u}_{2j}\right)
	\right],
\end{eqnarray}
where $\Delta$ is the two-dimensional Laplace operator, $\delta({\bf r})$ is the delta-function, and we take into account that $\sin({\bf k}_0{\bf u}_j)=\pm\sqrt3/2$. We use also in Eq.~\eqref{contlim} that Eqs.~\eqref{sys} read at ${\bf r}_i={\bf 0}$ as 
\begin{equation}
\label{zeroch}
{\cal J}\sqrt\beta\sum_{j=1}^6\rho_{{\bf u}_j}=0. 
\end{equation}
Then, the solution of Eq.~\eqref{contlim} is the electric field produced by six charges located at lattice sites around the defect (see Fig.~\ref{lattice}) whose total charge is zero (see Eq.~\eqref{zeroch}). The field of a single charge $q$ located at ${\bf r} = {\bf r}_0$ is given by the well known solution of the two-dimensional Poisson equation $\rho({\bf r})=\frac{q}{2\pi}\ln |{\bf r} - {\bf r}_0|$. Due to the linearity of the equation, the solution of Eq.~\eqref{contlim} is a sum of fields from six charges (the superposition principle in electrostatics)
\begin{eqnarray}
\label{solu1}
  \rho({\bf r})
	&=&
	\frac{U}{2\pi} \sqrt{\frac{2S}{3}}
	\sum_{j=1}^3
	\left[
	\left( \sqrt{\frac{2}{3S}}\rho_{{\bf u}_{2j-1}}+1 \right) \ln\left|{\bf r}-{\bf u}_{2j-1}\right|
	+
	\left( \sqrt{\frac{2}{3S}}\rho_{{\bf u}_{2j}}-1 \right) \ln\left|{\bf r}-{\bf u}_{2j}\right|
	\right].
\end{eqnarray}	

One expects that the solution of Eq.~\eqref{contlim} describes well the solution of Eqs.~\eqref{sys} not very close to the defect, near which $\rho({\bf r})$ may change rapidly and the transition to the continuum limit may be invalid. However, Eq.~\eqref{contlim} is not self-contained as it depends on $\rho_{{\bf u}_j}$. We find $\rho_{{\bf u}_j}$ by considering six equations \eqref{sys} for ${\bf r}_i={\bf u}_i$. But these six equations contain $\rho_{{\bf r}_i}$ from the second nearest neighbor sites. Assuming that solution \eqref{solu1} works at those sites, we come to a system of six linear equations on $\rho_{{\bf u}_i}$ presented in Appendix~\ref{rhoeq} which can be readily solved with the result
\begin{equation}
\label{solrho0}
\begin{aligned}
\rho_{{\bf u}_1} &= -\rho_{{\bf u}_2} = \rho_{{\bf u}_3} = -\rho_{{\bf u}_4} = \rho_{{\bf u}_5} = -\rho_{{\bf u}_6} = \sqrt{\frac{3S}{2}}\tilde\rho,\\
\tilde\rho &= -U\frac{3 \pi +\ln \frac{9}{7}}{3 \pi  (8+U)+U \ln\frac{9}{7}}.
\end{aligned}
\end{equation}
Notice that $U\tilde\rho\le0$ at $U>-7.8$ and $|\tilde\rho|\ll1$ at $|U|\alt2$. 
Eq.~\eqref{solu1} can be represented in the following more compact form far from the defect:
\begin{equation}
\label{quadrup}
	\rho({\bf r}) = -U\frac{1+\tilde\rho}{\pi} \sqrt{\frac{2S}{3}} \frac{\cos(3\phi)}{r^3}, \qquad r\gg1,
\end{equation}
where $\phi$ is the azimuthal angle of $\bf r$ counted from $x$ axis (see Fig.~\ref{lattice}). The octupolar pattern in Eq.~\eqref{quadrup} was found first analytically and numerically in Ref.~\cite{wollny1} for the vacancy. 
\footnote{
As it was suggested first in Ref.~\cite{wollny1}, values of additional turns of spins $\delta \theta (r)$ at distant $r\gg1$ from a single defect can be estimated in a $d$-dimensional system as
$
  \delta \theta (r) \sim	
  \int d^d{\bf q} e^{i{\bf qr}}\chi_\perp({\bf q},\omega=0)h_{\bf q},
$
where $\chi_\perp({\bf q},\omega=0)$ is the static transverse susceptibility and $h_{\bf q}$ is the Fourier transform of an effective local magnetic field acting on spins adjacent to the impurity. In AFs on the triangular lattice with defects of the considered type, $\chi_\perp({\bf q},\omega=0)\propto (\epsilon_{\bf q}^{(0)})^{-2}$, $h_{\bf q}\propto q^3$, and $\delta \theta (r) \sim 1/r^{d+1}$ at $r\gg1$ (cf.\ Eq.~\eqref{quadrup}). Then, one obtains $\delta \theta (r) \sim 1/r^{d-1}$ for a defect bond because $h_{\bf q}\propto q$ in this case (see also Ref.~\cite{olegdm}). 
}
It is clear from the above consideration that the distortion of the bare magnetic order made by a single-bond defect decays slowly ($\sim1/r$) because it is given by a field from one electric dipole (see also Ref.~\cite{olegdm}).

The total magnetic moment $\bf s$ of the spin texture produced by one defect and found from Eqs.~\eqref{transform}, \eqref{spinrep}, \eqref{shift}, and \eqref{phi} reads as
\begin{eqnarray}
\label{sx}
	s_x &=& \sqrt{2S} \sum_j \rho_{{\bf r}_j} \cos({\bf k}_0{{\bf r}_j}) = 0,\\
\label{sz}
	s_z &=& S\varepsilon_{{\bf k}_0}= -\sqrt{2S} \sum_j \rho_{{\bf r}_j} \sin({\bf k}_0{{\bf r}_j}) 
	\approx
	S \Bigl( 9\tilde\rho + 0.156 \left(1+\tilde\rho \right)U \Bigr),
\end{eqnarray}
where we use Eq.~\eqref{solrho0} for ${\bf r}_j={\bf u}_j$ and Eq.~\eqref{solu1} for the rest sites.

Interestingly, it follows from Eqs.~\eqref{u}, \eqref{solu1}, and \eqref{quadrup} that the bare $120^\circ$ magnetic order is not disturbed at $S'\ne S$ and ${\cal J}=S/S'\ne1$ (when $U=0$). It is quite natural because the energy per dashed bond $-{\cal J}SS'/2$ is equal to the energy per solid bond $-S^2/2$ in the $120^\circ$ order in this case (see Fig.~\ref{lattice}). However, we show in Sec.~\ref{subst} that the spin dynamics does change at ${\cal J}=S/S'\ne1$ because the energy costs differ of a deviation of the spin at the center of the defect and of other spins.

\underline{\it Vacancy}. In particular, we obtain $\tilde\rho \approx 0.147$ from Eqs.~\eqref{solrho0} in the case of the vacancy ($U=-1$). At this value of $\tilde\rho$, the angle of deviation of spins at ${\bf r}_i={\bf u}_i$ from their directions in the bare $120^\circ$ order is given by $\sqrt{3}\tilde\rho\approx0.255$ (see Eqs.~\eqref{theta} and \eqref{solrho0}) that is approximately 13\% as large as numerical findings of Ref.~\cite{wollny1} which are reproduced by our formulas at
\begin{equation}
\label{solrho}
\tilde\rho \approx 0.127.
\end{equation}
This agreement can be considered as a good one in view of approximations made in our analytical discussion.

The magnetic moment $\bf s$ produced by one vacancy in the system is given by Eqs.~\eqref{sx}, \eqref{sz}, and \eqref{solrho}, where now $S$ (the spin value at the center of the defect) should be subtracted from Eq.~\eqref{sz} so that $s_x=0$ and $s_z\approx -0.03S$. Then, the spin texture overcompensates the removed spin that is also in a good quantitative agreement with results of previous numerical simulations~\cite{wollny1}.

\subsection{Small concentration of defects}

It is seen from the above consideration that even the largest angle of deviation of spins due to defects ($\approx0.2$) is much smaller than unity at $|U|\alt1$. This fact justifies our assumption \eqref{assum} and our restriction to the leading orders in $\rho_{{\bf r}_j}$. Due to the linearity of all equations on $\rho_{{\bf r}_j}$, one can easily obtain the distortion of the bare magnetic order at small concentration $c$ of such defects as
\begin{equation}
\label{rhoc}
	\rho_{{\bf r}_j} = \sum_{{\bf r}_d} \rho^{(s)}_{{\bf r}_j-{\bf r}_d},
\end{equation}
where ${\bf r}_d$ determines defects positions and $\rho^{(s)}_{{\bf r}_i}$ is the single-impurity solution which is given by Eq.~\eqref{solrho0} for sites nearest to the defect center and by Eq.~\eqref{solu1} for the rest sites.

\subsection{Elastic neutron scattering cross section}
\label{ENS1}

We calculate now the cross-section of the elastic neutron scattering at small defect concentration $c\ll1$ which has the form \cite{Lowesey}
\begin{equation}
\label{cs}
  \frac{d \sigma}{d \Omega} \propto
	\sum_{i,j} e^{i {\bf Q}(\mathbf{r}_{i}-\mathbf{r}_{j})}
	\sum_{\chi,\eta} (\delta_{\chi\eta}-{\widehat{ Q}}_\chi\widehat{Q}_\eta)
	\langle S^\chi_{i}\rangle \langle S^\eta_{j}\rangle,
\end{equation}
where $\bf Q$ is the momentum transfer, $\widehat{\bf Q}={\bf Q}/Q$, $\chi,\eta=x,y,z$, $ \langle \dots \rangle$ denotes an average over quantum and thermal fluctuations. In our case
\begin{equation}
\label{sxz}
\begin{aligned}
	\langle S^{x}_{{\bf r}_j} \rangle &= S \sin({\bf k}_0{{\bf r}_j}) + \sqrt{2S} \rho_{{\bf r}_j} \cos({\bf k}_0{\bf r}_j) 
	- \rho^2_{{\bf r}_j} \sin({\bf k}_0{\bf r}_j),\\
	\langle S^{z}_{{\bf r}_j} \rangle &= S \cos({\bf k}_0{{\bf r}_j}) - \sqrt{2S} \rho_{{\bf r}_j} \sin({\bf k}_0{\bf r}_j) 
	- \rho^2_{{\bf r}_j} \cos({\bf k}_0{\bf r}_j),
\end{aligned}
\end{equation}
where we take into account only terms ${\cal O}(\rho^2_{{\bf r}_j})$ and $\rho_{{\bf r}_j}$ is given by Eq.~\eqref{rhoc}. Terms in Eqs.~\eqref{sxz} not containing $\rho_{{\bf r}_j}$ give the well known result for a triangular antiferromagnet without disorder
\begin{equation}
  \label{PureCS}
	  \left(\frac{d \sigma}{d \Omega}\right)_{\rm Bragg} \propto
  2 \pi^3 N S^2\left(1+\widehat{Q}_y^2\right) \sum_{\mbox{\boldmath $\tau$}}
  \left(\delta({\bf Q}+{\bf k}_0-\mbox{\boldmath $\tau$})+\delta({\bf Q}-{\bf k}_0-\mbox{\boldmath $\tau$})\right),
\end{equation}
where $N$ is the number of sites in the lattice, $\mbox{\boldmath $\tau$}$ are reciprocal lattice vectors, and delta-functions describe magnetic Bragg peaks at ${\bf Q}=\pm {\bf k}_0+\mbox{\boldmath $\tau$}$. Terms in Eq.~\eqref{cs} linear in $\rho_{{\bf r}_j}$ give zero after averaging over disorder configurations. Most of the quadratic in $\rho_{{\bf r}_j}$ terms give either zero or contributions proportional to Eq.~\eqref{PureCS} with a small factor $c$. The only important quadratic in $\rho_{{\bf r}_j}$ term, which determines the diffuse magnetic scattering cross section, has the form
\begin{eqnarray}
\label{diff}
\left( \frac{d \sigma}{d \Omega} \right)_{\rm diff} &=&
  S\left(1 + \widehat{Q}_y^2\right)
	\overline{
	\sum_{p,j} e^{i {\bf Q}\left(\mathbf{r}_{p}-\mathbf{r}_{j}\right)}
	\rho_{{\bf r}_p}\rho_{{\bf r}_j} \cos\left({\bf k}_0(\mathbf{r}_{p}-\mathbf{r}_{j})\right)
	}
	\approx
	c N S \left(1+\widehat{Q}_y^2\right) \frac12
	\left({\cal F}({\bf Q}+{\bf k}_0)
		+
	{\cal F}({\bf Q}-{\bf k}_0)
	\right),\\
	\label{f}
	{\cal F}({\bf k}) &=& \left|\sum_j \rho_{{\bf r}_j}^{(s)} e^{i{\bf k}{\bf r}_j}\right|^2,
\end{eqnarray}
where the line denotes the averaging over disorder configurations and $\rho_{{\bf r}_j}^{(s)}$ stands in Eq.~\eqref{f} for the single-defect solution. The summation in Eq.~\eqref{f} can be carried out analytically when $\bf k$ is close to a reciprocal lattice vector $\mbox{\boldmath $\tau$}$ in which case
$
\sum_j \rho_{{\bf r}_j} e^{i \mathbf{k} {{\bf r}_j}} 
\approx
\int d{\bf r}\rho({\bf r}) e^{i (\mathbf{k}-\mbox{\boldmath $\tau$}) \mathbf{r}}
$
and one obtains using Eqs.~\eqref{solu1} and \eqref{solrho0} 
\begin{equation}
  \label{fourier}
	{\cal F}({\bf k}) \approx \frac83 S U^2 (1+\tilde\rho)^2
	\frac{\left(\sum_{j=1}^3\sin{\left({\bf k}{\bf u}_{2j-1}\right)}\right)^2}{\left({\bf k}-{{\mbox{\boldmath $\tau$}}}\right)^4}.
\end{equation}
Notice that there is no singularity in Eq.~\eqref{fourier} at ${\bf k}={{\mbox{\boldmath $\tau$}}}$ because 
$
\sum_{j=1}^3\sin{\left({\bf k}{\bf u}_{2j-1}\right)} \sim |{\bf k}-{{\mbox{\boldmath $\tau$}}}|^3
$.
This should be contrasted with the single-bond disorder in 3D spiral magnets, where ${\cal F}({\bf k}) \sim 1/|{\bf k}-{{\mbox{\boldmath $\tau$}}}|^2$ and Bragg peaks acquire power-law decaying tails. \cite{olegdm}

\section{Magnon spectrum renormalization at $S'=S$}
\label{magnon}

We discuss now the impact of small concentration $c$ of defects on the magnon spectrum within the self-consistent $T$-matrix approach. To light notations, we assume in this section that the spin value at the center of the defect $S'$ is equal to the spin value in the host system $S$. In particular, results of this section is applicable to the vacancy. We consider arbitrary $S'\ne0$ in Sec.~\ref{subst} and show that expressions derived for $S'=S$ remain valid for $S'\ne S$ under simple replacements.

\begin{figure}
  \noindent
  \includegraphics[scale=0.8]{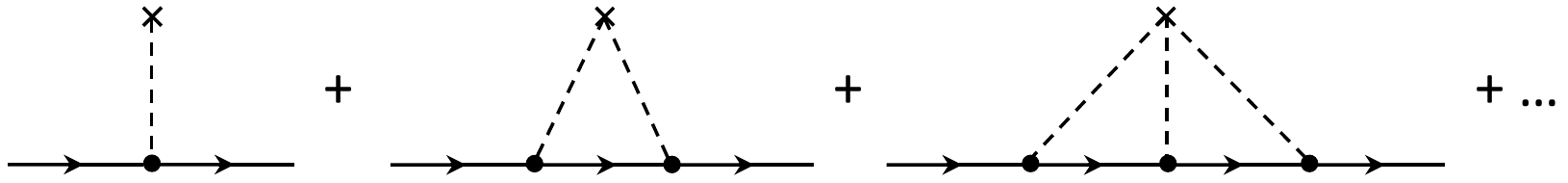}
  \caption{Diagrams for self-energy parts which should be taken into account in the first order in the defects concentration.}
  \label{diag}
\end{figure}

One has to calculate diagrams shown in Fig.~\ref{diag} to find self-energy parts in the first order in $c$. Only terms bilinear in Bose operators contribute to these diagrams arisen due to the disorder. Then, we have to take into account Eq.~\eqref{v2} and the bilinear terms appearing in Eqs.~\eqref{h3} and \eqref{v3} after shift \eqref{shift} (i.e., we take into account terms ${\cal O}(\rho_{{\bf r}_j})$). Thus, we obtain from Eqs.~\eqref{v2} and \eqref{v3}
\begin{eqnarray}
\label{v2t}
	\widetilde{\mathcal{V}}_2 &=&
	U\frac{S}{4}\sum_{{\bf r}_d}\sum_{j=1}^6
	\Bigl( (2-6\tilde\rho)a^\dagger_{{\bf r}_d} a^{}_{{\bf r}_d} 
	+ (2-3\tilde\rho)a^\dagger_{{\bf r}_d+{\bf u}_j}a^{}_{{\bf r}_d+{\bf u}_j} 
	+(1+3\tilde\rho)
	\left( a^\dagger_{{\bf r}_d}a^{}_{{\bf r}_d+{\bf u}_j} 
	+a^\dagger_{{\bf r}_d+{\bf u}_j}a^{}_{{\bf r}_d} \right) 
	\nonumber\\
	&&
	+(3\tilde\rho-3)
	\left( a^\dagger_{{\bf r}_d+{\bf u}_j}a^\dagger_{{\bf r}_d} 
	+a^{}_{{\bf r}_d+{\bf u}_j}a^{}_{{\bf r}_d} \right)
	-\frac{3\tilde\rho}{4} 
	\left( a^\dagger_{{\bf r}_d+{\bf u}_j}a^\dagger_{{\bf r}_d+{\bf u}_j} 
	+ a^{}_{{\bf r}_d+{\bf u}_j}a^{}_{{\bf r}_d+{\bf u}_j}\right) 
	\Bigr),
\end{eqnarray}
where $\tilde\rho$ is introduced in Eq.~\eqref{solrho0} (e.g., $\tilde\rho$ is given by Eq.~\eqref{solrho} for the vacancy at $U=-1$) and ${\bf r}_d$ determines positions of defects in the lattice. Eq.~\eqref{h3} gives
\begin{equation}
\label{h3t}
	\widetilde{\mathcal{H}}_2 = 
	\sqrt{\frac S2} \sum_i\sum_{j=1}^6 
	\rho_{{\bf r}_i}
	\sin({\bf k}_0{\bf u}_j)
	\left(
	\left( a^\dagger_{{\bf r}_i} + a^{}_{{\bf r}_i} \right)
	\left(a^\dagger_{{\bf r}_i+{\bf u}_j} + a^{}_{{\bf r}_i+{\bf u}_j} \right)
	-
	2 a^\dagger_{{\bf r}_i+{\bf u}_j}  a^{}_{{\bf r}_i+{\bf u}_j}
	\right),
\end{equation}
where $\rho_{{\bf r}_i}$ is given by Eq.~\eqref{rhoc}.

Solid lines in diagrams shown in Fig.~\ref{diag} contain sums over momenta of the form $\sum_{\bf q}{\cal G}(\omega,{\bf q})$, where ${\cal G}$ stands for one of the Green's functions \eqref{gf}. Such sums show logarithmic singularities at $\omega^2-\Omega(\omega,{\bf q})\to0$ stemming from the linear dispersion of the bare spectrum near $\bf 0$ and $\pm{\bf k}_0$ (see Eqs.~\eqref{d} and \eqref{spec0l}). However, there are "soft" combinations of Green's functions which do not have such singularities due to additional smallness in numerator. For instance, $F(\omega,{\bf q})+G(\omega,{\bf q})$ contains only terms in numerator ${\cal O}(({\bf q}\mp{\bf k}_0)^2)$ and ${\cal O}(\omega)$ at ${\bf q}\approx \pm{\bf k}_0$, and there are only terms ${\cal O}(q^2)$ and ${\cal O}(\omega)$ in numerator of $F(\omega,{\bf q})-G(\omega,{\bf q})$ at $q\ll1$. Besides, some contributions from $\widetilde{\mathcal{H}}_2$ (see Eq.~\eqref{h3t}) can be omitted because the Fourier transform of $\rho_{{\bf r}_i}$ is ${\cal O}(q)$ at $q\ll1$ (see Eqs.~\eqref{f} and \eqref{fourier}).  We take into account below only the most singular contribution to each diagram. 

\subsection{$k\ll1$}

We proceed with the discussion of the spectrum renormalization at $k\ll1$ in the first order in $c$. It is seen from Eqs.~\eqref{h2f} and \eqref{o} that
\begin{equation}
\label{oll1}
\Omega(\omega,{\bf k}) \approx \frac{9S}{2}\left(\Sigma + \overline{\Sigma} + \Pi + \Pi^\dagger\right)
\end{equation}
at $k\ll1$, where we neglect terms ${\cal O}(ck^2)$, we omit $- \Pi\Pi^\dagger + \Sigma \overline{\Sigma}$ because it gives corrections ${\cal O}(c^2)$, and we discard also $\omega (\Sigma - \overline{\Sigma})$ as it contributes to the renormalization of the Green's functions residues and gives corrections to the spectrum square ${\cal O}(c^2)$ and ${\cal O}(ck^2)$. Thorough analysis shows that the most singular contributions to Eq.~\eqref{oll1} arise from small momenta of summation $q\ll1$ and one comes to a simple geometric series. We obtain as a result the following equation on $\Omega(\omega,{\bf k})$ within the self-consistent $T$-matrix approach:
\begin{equation}
\label{eqo}
\Omega(\omega,{\bf k}) = c 
\frac{3 v_\|^2U\tilde\rho}{1-\frac{3}{4\pi} U\tilde\rho \ln(-(\omega^2-\Omega(\omega,{\bf k}))/v_\|^2)},
\end{equation}
where the logarithm in the denominator comes from 
$
\sum_{\bf q}(\omega^2-v_\|^2q^2-\Omega(\omega,{\bf q}))^{-1}
$
and we take into account that all the most singular diagrams for $\Omega(\omega,{\bf k})$ do not depend on $\bf k$ (i.e., $\Omega(\omega,{\bf q})=\Omega(\omega,{\bf k})$). 
One does not need to solve Eq.~\eqref{eqo} in order to find the spectrum $\epsilon_{\bf k}$ which is determined by the equation (see Eq.~\eqref{d})
\begin{equation}
\label{speceq}
{\cal D}(\epsilon_{\bf k},{\bf k})  = \epsilon_{\bf k}^2 - v_\|^2k^2 - \Omega(\epsilon_{\bf k},{\bf k})	= 0.
\end{equation}
It is seen from Eq.~\eqref{speceq} that $\epsilon_{\bf k}^2 - \Omega(\epsilon_{\bf k},{\bf k})= v_\|^2k^2$. Substituting this equality to Eq.~\eqref{eqo} at $\omega=\epsilon_{\bf k}$ and coming back to Eq.~\eqref{speceq}, we obtain for the spectrum square
\begin{equation}
\label{specr1}
	\epsilon_{\bf k}^2 = v_\|^2k^2 + 
	c \frac{3v_\|^2U\tilde\rho }{1 + \frac{3}{4\pi} U\tilde\rho \left(-2\ln k+i\pi\right)}.
\end{equation}
It is convenient to introduce the following characteristic momenta for further analysis of Eq.~\eqref{specr1}:
\begin{eqnarray}
\label{kappa1}
\kappa_1 &=& \sqrt{ 3|U\tilde\rho|c },\\
\kappa_2 &=& \exp\left( -\frac{2\pi}{3|U\tilde\rho|} \right).
\end{eqnarray}
One obtains from  Eq.~\eqref{specr1} at $\kappa_1\gg\kappa_2$
\begin{equation}
\label{resspec}
\epsilon_{\bf k} \approx \left\{
\begin{aligned}
	& v_\|k + c\frac{v_\|}{k}\frac{3}{2}U\tilde\rho - ic\frac{v_\|}{k}\frac{9}{8}\left(U\tilde\rho\right)^2, \quad & \kappa_1\ll k \ll1, & \quad {\rm (i)}\\
	& v_\|\sqrt{3 U \tilde\rho c} \left( 1-i\frac{3}{8}U\tilde\rho \right), \quad & \kappa_2\ll k \ll \kappa_1, & \quad {\rm (ii)}\\
	& v_\|\sqrt{ \frac{2\pi c}{-\ln k} } \left( 1 + i \frac{\pi}{4\ln k} \right), \quad & k \ll \kappa_2, & \quad {\rm (iii)}
\end{aligned}
\right.
\end{equation}
where the first terms in all regimes are much larger than the rest terms and imaginary values represent the magnon damping. 

\begin{figure}
  \noindent
  \includegraphics[scale=1]{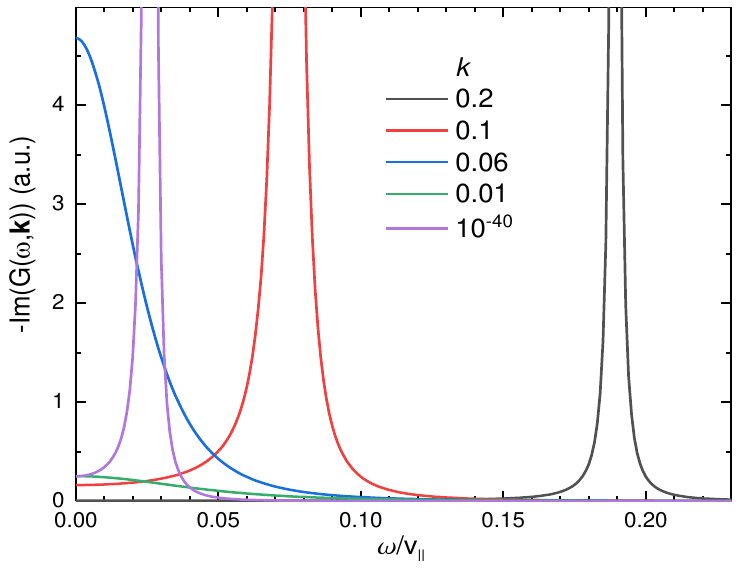}
  \caption{Imaginary part of the Green's function $G(\omega,{\bf k})$ (see Eq.~\eqref{gf}) at some representative momenta $k\ll1$ calculated numerically using Eqs.~\eqref{d} and \eqref{eqo} for vacancies ($U=-1$) at $c=0.01$. Well defined magnons are seen at $k\gg\kappa_1\approx0.06$ and $k\ll\kappa_2\sim10^{-8}$ whereas magnons are overdamped at $\kappa_2\ll k\ll\kappa_1$ in agreement with analytical finding \eqref{resspec}.
	}
  \label{dsf}
\end{figure}

It is seen from Eq.~\eqref{resspec} that the bare spectrum is renormalized drastically at $k\ll\kappa_1\sim\sqrt c$. Notice that the real part of the spectrum is much smaller than the imaginary one in regime (ii) at both positive and negative $U$ because $U\tilde\rho\le0$. As a result, a broad asymmetric anomaly arises in the dynamical structure factor at corresponding $\bf k$ which is normally attributed to the magnon localization. \cite{oleg,2dvac,sheng} Our estimations show that regime (iii) in Eq.~\eqref{resspec} is realized normally at extremely small momenta (e.g., $\kappa_2\sim 10^{-8}$ for the vacancy). Fig.~\ref{dsf} illustrates these findings in the particular case of vacancies at $c=0.01$. It is demonstrated below that the appearance of a broad range of momenta (and energies) governed by the localized states is reflected in the density of states and in the specific heat in a wide range of small temperatures.

The counterpart of Eq.~\eqref{resspec} can be readily found from Eq.~\eqref{specr1} in the opposite limiting case of $\kappa_1\ll\kappa_2$. Because this regime is realized at extremely small concentrations (e.g., at $c\ll 10^{-14}$ for the vacancy), we point out only that the bare spectrum $\epsilon_{\bf k} \approx v_\|k$ is renormalized weakly at $k\gg \kappa_1\sim\sqrt c$, whereas $\epsilon_{\bf k}$ is given by Eq.~\eqref{resspec} in regime (iii) at $k\ll \kappa_1$.

\subsection{${\bf k}\approx \pm {\bf k}_0$}

Similarly to the case of $k\ll1$, one obtains two geometric series for 
$\Omega(\omega,{\bf k}) \approx \frac{9S}{4}\left(\Sigma + \overline{\Sigma} - \Pi - \Pi^\dagger\right)$ 
at $|{\bf k} \pm {\bf k}_0|\ll1$, and the counterpart of Eq.~\eqref{specr1} reads as
\begin{equation}
\label{specr1k0}
	\epsilon_{\bf k}^2 = v_\perp^2|{\bf k} \pm {\bf k}_0|^2 
	+ c v_\perp^2
	\left(
	\frac{Q_1}{1+\frac{Q_1}{2\pi} \left(-2\ln |{\bf k} \pm {\bf k}_0|+i\pi\right)}
	+
	\frac{Q_2}{1+\frac{Q_2}{2\pi} \left(-2\ln |{\bf k} \pm {\bf k}_0|+i\pi\right)}
	\right),
\end{equation}
where
\begin{equation}
\label{qs}
	\begin{aligned}
	Q_1 &= -3 \left( U+\varepsilon_{{\bf k}_0} \right) - \frac{51}{4} U\tilde\rho, \\
	Q_2 &= 3 \left( U+\varepsilon_{{\bf k}_0} \right) - \frac{9}{4} U\tilde\rho,
	\end{aligned}
\end{equation}
$\varepsilon_{{\bf k}_0}$ is defined in Eq.~\eqref{sz} and it is related to the Fourier transform of the single-impurity solution $\rho_{{\bf r}_i}^{(s)}$ at ${\bf k}={\bf k}_0$. As a result, one obtains from  Eq.~\eqref{specr1k0} at $\kappa_1\gg\kappa_2$ (cf.\ Eq.~\eqref{resspec})
\begin{equation}
\label{resspeck0}
\epsilon_{\bf k} \approx \left\{
\begin{aligned}
	& v_\perp |{\bf k} \pm {\bf k}_0| 
	+ c\frac{v_\perp \left(Q_1+Q_2\right)}{2|{\bf k} \pm {\bf k}_0|} 
	- ic\frac{v_\perp \left(Q_1^2+Q_2^2\right)}{4|{\bf k} \pm {\bf k}_0|}, \quad & \kappa_1\ll |{\bf k} \pm {\bf k}_0| \ll1, & \quad {\rm (i)}\\
	& v_\perp \sqrt{c \left( Q_1+Q_2 - i \frac{Q_1^2+Q_2^2}{2} \right)}, \quad & \kappa_2\ll |{\bf k} \pm {\bf k}_0| \ll \kappa_1, & \quad {\rm (ii)}\\
	& v_\perp\sqrt{ \frac{\pi c}{-\ln |{\bf k} \pm {\bf k}_0|} } \left( 1 + i \frac{\pi}{4\ln |{\bf k} \pm {\bf k}_0|} \right), \quad & \kappa_3\ll|{\bf k} \pm {\bf k}_0| \ll \kappa_2, & \quad {\rm (iii)}\\
	& v_\perp\sqrt{ \frac{2\pi c}{-\ln |{\bf k} \pm {\bf k}_0|} } \left( 1 + i \frac{\pi}{4\ln |{\bf k} \pm {\bf k}_0|} \right), \quad & |{\bf k} \pm {\bf k}_0| \ll \kappa_3, & \quad {\rm (iv)}
\end{aligned}
\right.
\end{equation}
where the first terms in regimes (i), (iii), and (iv) are much larger than the rest terms, imaginary values represent the magnon damping, $\kappa_1$ is given by Eq.~\eqref{kappa1}, and
\begin{equation}
	\begin{aligned}
	\kappa_2 &= \exp\left( -\frac{\pi}{\max\{|Q_1|,|Q_2|\}} \right),\\
	\kappa_3 &= \exp\left( -\frac{\pi}{\min\{|Q_1|,|Q_2|\}} \right).
	\end{aligned}
\end{equation}
Particular estimations show that $Q_1+Q_2>0$ and $Q_1+Q_2\gg Q_1^2+Q_2^2$ at $|U|\alt0.6$. Then, magnons are well-defined quasiparticles with practically dispersionless spectrum in regime (ii) at $|U|\alt0.6$. In contrast, $Q_1+Q_2\sim Q_1^2+Q_2^2$ in regime (ii) at $1\sim|U|\agt0.6$ and magnons are overdamped.

In contrast to the case of $k\ll1$, regime (iii) in Eq.~\eqref{resspeck0} can be achieved at not too small momenta (e.g., $\kappa_2\approx0.17$ for the vacancy). This signifies also that $\kappa_1$ can be smaller than $\kappa_2$ at reasonably small concentrations and not too small $|U\tilde\rho|$ (e.g., at $c<0.07$ for the vacancy). Then, we find at $\kappa_1\ll\kappa_2$
\begin{equation}
\label{resspeck02}
\epsilon_{\bf k} \approx \left\{
\begin{aligned}
	& v_\perp |{\bf k} \pm {\bf k}_0| 
	+ c\frac{v_\perp \left(Q_1+Q_2\right)}{2|{\bf k} \pm {\bf k}_0|} 
	- ic\frac{v_\perp \left(Q_1^2+Q_2^2\right)}{4|{\bf k} \pm {\bf k}_0|}, \quad & \kappa_2\ll |{\bf k} \pm {\bf k}_0| \ll1, & \quad {\rm (i)}\\
	& v_\perp |{\bf k} \pm {\bf k}_0| 
	+ v_\perp\sqrt{ \frac{\pi c}{-\ln |{\bf k} \pm {\bf k}_0|} } \left( 1 + i \frac{\pi}{4\ln |{\bf k} \pm {\bf k}_0|} \right), \quad & \kappa_1\ll |{\bf k} \pm {\bf k}_0| \ll \kappa_2, & \quad {\rm (ii)}\\
	& v_\perp\sqrt{ \frac{\pi c}{-\ln |{\bf k} \pm {\bf k}_0|} } \left( 1 + i \frac{\pi}{4\ln |{\bf k} \pm {\bf k}_0|} \right), \quad & \kappa_3\ll|{\bf k} \pm {\bf k}_0| \ll \kappa_1, & \quad {\rm (iii)}\\
	& v_\perp\sqrt{ \frac{2\pi c}{-\ln |{\bf k} \pm {\bf k}_0|} } \left( 1 + i \frac{\pi}{4\ln |{\bf k} \pm {\bf k}_0|} \right), \quad & |{\bf k} \pm {\bf k}_0| \ll \kappa_3, & \quad {\rm (iv)}
\end{aligned}
\right.
\end{equation}
where $\kappa_1=\sqrt{\pi c/\ln(\pi c)}$. Notice that the damping is much smaller than the magnon energy in all regimes in Eq.~\eqref{resspeck02} and the strong bare spectrum renormalization takes place at $|{\bf k} \pm {\bf k}_0|\ll\kappa_1\sim\sqrt c$. 

\section{Magnon spectrum renormalization at $S'\ne S$}
\label{subst}

We assume now that the spin value $S'$ at the center of the defect differs from the spin value in the host system $S$. It can be shown that results for the spectrum \eqref{kappa1}--\eqref{resspec} at $k\ll1$ remain valid under the replacement
\begin{equation}
\label{replace2}
U \tilde\rho \mapsto U \tilde\rho + \frac43 {\cal J} (1-3\tilde\rho)\left(\sqrt\beta-1\right)^2,
\end{equation}
where $\beta$ is given by Eq.~\eqref{beta}. Notice that magnons can be well-defined quasiparticles in regime (ii) because Eq.~\eqref{replace2} is positive in a certain range of parameters. The spectrum at ${\bf k}\approx \pm {\bf k}_0$ is given by Eqs.~\eqref{resspeck0} and \eqref{resspeck02}, where now (cf.\ Eqs.~\eqref{qs})
\begin{eqnarray}
\label{q1s}	
	Q_1 &=& -3 \left( U + \varepsilon_{{\bf k}_0} \right) - \frac{51}{4} U\tilde\rho
	+4 {\cal J}\left( \left( \sqrt \beta-1 \right)^2 + 3\tilde\rho (\beta-1) \right), \\
	Q_2 &=& 3 \left( U + \varepsilon_{{\bf k}_0} \right) - \frac{9}{4} U\tilde\rho.
\end{eqnarray}

Interestingly, even when the bare magnetic ordering is not disturbed at $U = \tilde\rho=\varepsilon_{{\bf k}_0}=0$ and ${\cal J}=S/S'\ne1$, the strong spectrum renormalization at $k\ll\sqrt c$ and $|{\bf k}\pm{\bf k}_0|\ll\sqrt c$ discussed above does arise due to the finiteness of the last terms in Eqs.~\eqref{replace2} and \eqref{q1s}. It is quite natural that the spin dynamics changes in this case because the energy costs differ at ${\cal J}\ne1$ of the deviation of the substitutional spins and spins in the host system.

\section{Static properties}
\label{static}

\subsection{Order parameter at $T=0$}

The mean order parameter can be found at $T=0$ in the leading order in $1/S$ from Eqs.~\eqref{spinrep} and \eqref{shift} as follows:
\begin{equation}
\label{m}
	m 
	=	\frac1N \overline{\sum_i \langle S^{z^{\prime}}_{{\bf r}_i}\rangle}
	= S - \frac1N \overline{\sum_i \left( \rho_{{\bf r}_i} \right)^2 }
	= S - c \sum_i \left( \rho_{{\bf r}_i}^{(s)} \right)^2
	= S \left( 1 - c \left( 9\tilde\rho^2 + 0.008\left( 1 + \tilde\rho \right)^2 U^2 \right) \right), 
\end{equation}
where $N$ is the number of sites in the lattice, the line denotes the averaging over disorder configurations, $\rho_{{\bf r}_i}$ is given by Eq.~\eqref{rhoc}, and $\rho_{{\bf r}_i}^{(s)}$ stands for the single-defect solution. Notice that the sum in the right-hand side of Eq.~\eqref{m} converges at large $r_i$ (see Eq.~\eqref{quadrup}) whereas this sum diverges logarithmically for the single-bond defect in which case $\rho_{{\bf r}_i}\sim1/{r_i}$. \cite{olegdm} This divergence signifies a destruction of the long-range order by any finite concentration of single-bond defects in 2D systems. \cite{wollny3} In particular, we obtain from Eqs.~\eqref{m} and \eqref{solrho} for the vacancy $m=S(1-0.155c)$.

One has to discuss quantum corrections to coefficients in Eqs.~\eqref{sys} in order to find quantum renormalization of $m$. This consideration is out of the scope of the present paper.

\subsection{Critical temperature}

The strong renormalization of the long-wavelength magnon spectrum at finite $c\ll1$ leads to the stabilization of the slightly distorted $120^\circ$ long-range magnetic order at finite temperature $T<T_N$. One obtains for $T_N$ as a result of a standard estimation (which is based on the equality 
$
\frac1N \overline{\sum_i \langle S^{z^{\prime}}_{{\bf r}_i}\rangle} 
\approx 
m - \frac1N \overline{\sum_i \langle a^\dagger_{{\bf r}_i} a_{{\bf r}_i} \rangle} = 0
$, see Eq.~\eqref{m})
\begin{equation}
\label{tn}
T_N \sim S^2J \frac{1-Bc}{\ln(1/\kappa_1)} \sim S^2J \frac{1-Bc}{\ln(1/c)},
\end{equation}
where $B$ is a constant. One has to consider temperature corrections to coefficients in Eqs.~\eqref{sys} in order to find $B$ accurately in Eq.~\eqref{tn}. This consideration is out of the scope of the present paper.

\subsection{Density of states}

The density of states (DOS) $N(\omega)$ is expressed via the magnon retarded Green's function $\langle \mathfrak{b}_{\bf k}, \mathfrak{b}^\dagger_{\bf k} \rangle_\omega$ as follows (see Eqs.~\eqref{gf}--\eqref{o}):
\begin{equation}
\label{dens}
N(\omega) 
= -\frac1\pi {\rm Im} \sum_{\bf k} \langle \mathfrak{b}_{\bf k}, \mathfrak{b}^\dagger_{\bf k} \rangle_\omega
\approx -\frac1\pi {\rm Im} \sum_{\bf k} 
\frac{1}{2 \epsilon_{\bf k}^{(0)}} 
\frac{\left(\omega+\epsilon_{\bf k}^{(0)}\right)^2 + \left(\omega+\epsilon_{\bf k}^{(0)}\right) \left(\Sigma - \overline{\Sigma}\right) }{{\cal D}(\omega,{\bf k})},
\end{equation}
where $\mathfrak{b}_{\bf k}$ and $\mathfrak{b}^\dagger_{\bf k}$ are Bose operators related with $a_{\bf k}$ and $a^\dagger_{\bf k}$ by the Bogoliubov transformation which diagonalizes the bilinear part of the Hamiltonian \eqref{h2f}.

At $\omega\gg v_\| \kappa_1$, $N(\omega)\propto\omega$ is determined by weakly renormalized propagating short-wavelength spin waves (as in the pure system).

An accurate analysis of Eqs.~\eqref{d}, \eqref{eqo}, and \eqref{dens} shows that the overdamped magnons observed above (see regimes (ii) in Eqs.~\eqref{resspec} and \eqref{resspeck0}) give at $v_\| \kappa_2\ll\omega\ll v_\| \kappa_1$ a quadratic contribution to the DOS $N(\omega)\propto\omega^2/\sqrt c \ll\omega$.

Well-defined quasiparticles with the spectrum proportional to $\sqrt{c/\ln(1/k)}$ (see regime (iii) in Eq.~\eqref{resspec} and regimes (iii) and (iv) in Eqs.~\eqref{resspeck0} and \eqref{resspeck02}) give an exponentially small contribution to the DOS at $\omega\ll v_\| \kappa_2$: $N(\omega)\propto \exp(-4\pi cv^2/\omega^2)$.

Well-defined magnons with the dispersionless spectrum, which can arise in regimes (ii) in Eqs.~\eqref{resspec} and \eqref{resspeck0} in some range of parameters, produce a sharp peak (a delta-function, if one neglects the damping) at the corresponding energy.

\subsection{Specific heat}

Renormalization of the DOS discussed above is reflected in the magnetic part of the specific heat which is given by
\begin{equation}
\label{c}
C_m = \frac{1}{T^2} \int_0^\infty d\omega N(\omega)\omega^2
\left( \frac{1}{e^{\omega/T}-1} + \left(\frac{1}{e^{\omega/T}-1} \right)^2 \right).
\end{equation}
Eq.~\eqref{c} gives $C_m\propto T^2$ in the pure system and in the disordered one at 
$T \gg v_\|\kappa_1 \sim v_\|\sqrt c$. At smaller temperatures, this quadratic decreasing with $T$ changes to more faster ones: the contribution of the overdamped modes reads as $C_m\propto T^3/\sqrt c$ (at $v_\| \kappa_2\ll T\ll v_\| \kappa_1$) and magnons with the spectrum proportional to $\sqrt{c/\ln(1/k)}$ produce the exponential falling down of $C_m$.

Well-defined magnons with the dispersionless spectrum give an almost constant contribution to $C_m$ at $T$ larger than the magnon energy.

\begin{figure}
  \noindent
  \includegraphics[scale=1]{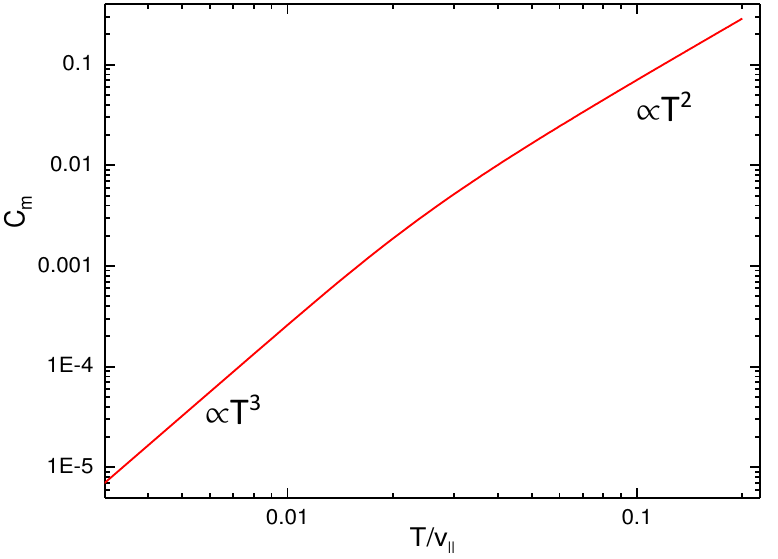}
  \caption{Specific heat $C_m$ for vacancies ($U=-1$) at $c=0.01$ calculated using Eq.~\eqref{c}. A crossover is seen from the pure-system behavior $C_m\propto T^2$ at $T\gg\kappa_1\approx0.06$ to the dependence $C_m\propto T^3$ at $T\ll\kappa_1$ due to the disorder-induced overdamped (localized) modes.
	}
  \label{heat}
\end{figure}

In particular case of the vacancy, the dependence $C_m\propto T^2$ at $T\gg\sqrt c$ is changed by $C_m\propto T^3/\sqrt c$ at $T\ll v_\|\sqrt c$ due to the overdamped modes (see Fig.\ref{heat} for illustration).

\section{Discussion}
\label{disc}

We point out that introduction to Hamiltonian \eqref{ham} of easy-plane anisotropy 
$
-A\sum_{\langle i,j \rangle} S^y_{{\bf r}_i} S^y_{{\bf r}_j}
$ 
produces a gap in the bare spectrum at ${\bf k}=\pm{\bf k}_0$. However our results of Sec.~\ref{order} about the distortion of the magnetic order, Eq.~\eqref{m} for the order parameter, Eq.~\eqref{tn} for the critical temperature, and Eq.~\eqref{resspec} for the spectrum renormalization at $k\ll1$ remain quantitatively valid in the $XY$ AF, where now $v_\|=\frac32 \sqrt{3-2A}S$. If $S'\ne S$, Eq.~\eqref{resspec} is valid under replacement \eqref{replace2}.

When some small low-symmetry spin interaction exists in the system which produces a gap $\Delta$ in the bare long-wavelength spectrum (e.g., an easy-axis anisotropy), our results are valid at $k\gg\Delta$ and $|{\bf k} \pm {\bf k}_0|\gg\Delta$. As it is shown in Ref.~\cite{oleg}, states are localized near the bottom of the excitation band in disordered systems with the bare spectrum of the form $\epsilon_{\bf k}=k^2+\Delta$.

In a quasi-2D system on the stacked triangular lattice and a small non-frustrated exchange interaction $J_\perp$ ($|J_\perp|\ll1$) between spins from neighboring triangular planes, our results for the spectrum renormalization are valid at $k\gg\sqrt{|J_\perp|}$ and $|{\bf k} \pm {\bf k}_0|\gg\sqrt{|J_\perp|}$. At $k\ll\sqrt{|J_\perp|}$ and $|{\bf k} \pm {\bf k}_0|\ll\sqrt{|J_\perp|}$, one has to replace $\ln k$ and $\ln|{\bf k}\pm{\bf k}_0|$ by $\ln\sqrt{|J_\perp|}$ in Eqs.~\eqref{specr1} and \eqref{specr1k0} and find counterparts of Eqs.~\eqref{resspec}, \eqref{resspeck0}, and \eqref{resspeck02} as a result of a simple analysis similar to that performed above. We do not present here the corresponding results and point out only that regimes with overdamped modes remain.

Renormalization of the magnon spectrum obtained above is much more stronger than that in the collinear 2D magnets found before \cite{wan,2dvac}. Technically, this stems from the fact that "soft" combinations of Green's functions appear in every diagram for $\Omega(\omega,{\bf k})$ in collinear magnets so that all diagrams are proportional to some powers of $k$ and $\omega$. In contrast, there are no such "soft" combinations in the most singular contributions to $\Omega(\omega,{\bf k})$ in non-collinear spin systems (see also Ref.~\cite{olegdm}).

\section{Conclusion}
\label{conc}

To conclude, we discuss the distortion of the long-range $120^\circ$ magnetic order and renormalization of the magnon spectrum in $XY$ and Heisenberg antiferromagnets on the triangular lattice with random vacancies and substitutional spins. The distortion from a single defect is described by equations for the field of electrically neutral complex of six charges surrounding the defect (see Fig.~\ref{lattice}) which is given by Eqs.~\eqref{u}, \eqref{solu1}, \eqref{solrho0}, and \eqref{quadrup}. In contrast to the case of the single-bond disorder, a finite concentration of such symmetric defects does not destroy the long-range magnetic order at $T=0$ and produces just a weak diffuse elastic neutron scattering.

We find that the spectrum of magnons whose wavelength is larger than the mean distance between impurities $1/\sqrt c$ (i.e., $k\ll\sqrt c$ and $|{\bf k} \pm {\bf k}_0| \ll \sqrt c$) is strongly renormalized due to the scattering on defects. At all considered model parameters, magnons are well-defined propagating quasiparticles at sufficiently small momenta whose spectrum is proportional to $\sqrt{c/\ln(1/k)}$ (see Eqs.~\eqref{resspec}, \eqref{resspeck0}, and \eqref{resspeck02}). At not too small momenta, we observe a broad range of model parameters (including the vacancy) at which magnons are overdamped and localized. The overdamped modes arise also in quasi-2D spin systems on the stacked triangular lattice. 

This strong renormalization of the long-wavelength spectrum leads to the stabilization of the slightly distorted long-range magnetic order at finite $T$ (the order-by-disorder effect) with the critical temperature given by Eq.~\eqref{tn}. The spectrum modification is reflected also in the low-$T$ behavior of the specific heat $C_m$. In particular case of the vacancy, the dependence $C_m\propto T^2$ in the pure system (and in the disordered one at $T\gg\sqrt c$) is changed by $C_m\propto T^3/\sqrt c$ at $T\ll\sqrt c$ due to the overdamped modes.

\begin{acknowledgments}

This work is supported by the Foundation for the Advancement of Theoretical Physics and Mathematics "BASIS".

\end{acknowledgments}

\appendix

\section{Equations for $\rho_{{\bf u}_i}$}
\label{rhoeq}

The set of equations for $\rho_{{\bf u}_i}$ has the form
\begin{eqnarray}
\label{rhoeqapp}
&&
-\rho_1 \left(6+U-U \frac{\ln 4}{6 \pi }\right) 
+ \rho_2 \left(1+U \frac{\ln3}{6 \pi }\right)
+\rho_3 U \frac{\ln28}{6 \pi }
+\rho_4 U \frac{\ln252}{6 \pi }
+\rho_5 U \frac{\ln7}{2 \pi }
+\rho_6 \left( 1 + U \frac{\ln84}{6 \pi }\right)
= U \left(1 +\frac{\ln\frac{9}{7}}{3 \pi }\right),\nonumber\\
&&
\rho_1 \left(1+U \frac{\ln12}{6 \pi }\right)
-\rho_2 (6+U)
+\rho_3 \left(1+U \frac{\ln12}{6 \pi }\right)
+\rho_4 U \frac{\ln14}{3 \pi }
+\rho_5 U \frac{\ln21}{3 \pi }
+\rho_6 U \frac{\ln14}{3 \pi }
=-U \left(1 +\frac{\ln\frac{9}{7}}{3 \pi }\right),\nonumber\\
&&
\rho_1 U \frac{\ln4}{6 \pi }
+\rho_2 \left(1+U \frac{\ln3}{6 \pi }\right)
-\rho_3 \left(6+U-U \frac{\ln28}{6 \pi }\right)
+\rho_4 \left(1+U \frac{\ln252}{6 \pi }\right)
+\rho_5 U \frac{\ln7}{2 \pi }
+\rho_6 U \frac{\ln84}{6 \pi }
=U \left(1 +\frac{\ln\frac{9}{7}}{3 \pi }\right),\nonumber\\
&&
\rho_1 U \frac{\ln12}{6 \pi }
+\rho_3 \left(1+U \frac{\ln12}{6 \pi }\right)
-\rho_4 \left(6+U-U \frac{\ln14}{3 \pi }\right)
+\rho_5 \left(1+U \frac{\ln21}{3 \pi }\right)
+\rho_6 U \frac{\ln14}{3 \pi }
=-U \left(1 +\frac{\ln\frac{9}{7}}{3 \pi }\right),\\
&&
\rho_1 U \frac{\ln4}{6 \pi }
+\rho_2 U \frac{\ln3}{6 \pi }
+\rho_3 U \frac{\ln28}{6 \pi }
+\rho_4 \left(1+U \frac{\ln252}{6 \pi }\right)
-\rho_5 \left(6+U-U \frac{\ln7}{2 \pi }\right)
+\rho_6 \left(1+U \frac{\ln84}{6 \pi }\right)
=U \left(1 +\frac{\ln\frac{9}{7}}{3 \pi }\right),\nonumber\\
&&
\rho_1 \left(1+U \frac{\ln12}{6 \pi }\right)
+\rho_3 U \frac{\ln12}{6 \pi }
+\rho_4 U \frac{\ln14}{3 \pi }
+\rho_5 \left(1+U \frac{\ln21}{3 \pi }\right)
-\rho_6 \left(6+U-U \frac{\ln14}{3 \pi }\right)
=-U \left(1 +\frac{\ln\frac{9}{7}}{3 \pi }\right),	\nonumber
\end{eqnarray}
where $\rho_i = \rho_{{\bf u}_i} \sqrt{\frac{2}{3S}}$. The solution of these equations is given by Eq.~\eqref{solrho0}.

\bibliography{vacancybib}

\end{document}